\documentclass[12pt]{article}
\usepackage{amsmath,latexsym,mathrsfs, amssymb}

\usepackage{color}
\usepackage{siunitx}
\usepackage{times}
\usepackage{lineno}

\usepackage[latin1]{inputenc}
\usepackage[pdftex]{graphicx}
\usepackage[labelfont=bf,justification=justified,   format=plain]{caption}
\DeclareGraphicsExtensions{.jpg,.pdf,.mps,.png,.ai}

\newenvironment{sciabstract}{%
\begin{quote} \bf}
{\end{quote}}

\title{Experimental evidence for long-distance electrodynamic intermolecular forces}

\author
{Mathias Lechelon,$^{1,2}$ Yoann Meriguet,$^{3,4}$ Matteo Gori,$^{1,5}$ Sandra Ruffenach,$^{4}$ \\
Ilaria Nardecchia,$^{1,2}$ Elena Floriani,$^{1}$ 
Dominique Coquillat,$^{4}$ \\ Fr\'ed\'eric Teppe,$^{4}$ S\'ebastien Mailfert,$^{2}$ \\
Didier Marguet,$^{2}$ Pierre Ferrier,$^{2}$ Luca Varani,$^{3}$\\
James Sturgis,$^{6}$ Jeremie Torres,$^{3,\ast}$ Marco Pettini,$^{1,\ast}$ \\
\\
\normalsize{$^{1}$ Aix-Marseille Univ, Universit\'e de Toulon, CNRS, Marseille, France  }\\
\normalsize{Centre de Physique Th\'eorique, CNRS, Marseille, France}\\
\normalsize{$^{2}$ Centre d'Immunologie de Marseille-Luminy, Aix-Marseille Universit\'e, } \\
\normalsize{CNRS, Inserm, Marseille, France}\\
\normalsize{$^{3}$ Institut d'Electronique et des Syst\`emes, University of Montpellier - CNRS, Montpellier, France } \\
\normalsize{$^{4}$ Laboratoire Charles Coulomb, University of Montpellier - CNRS,  Montpellier, France} \\
\normalsize{$^{5}$ Quantum Biology Lab, Howard University, 2400 6th St NW, Washington, DC 20059, USA} \\
\normalsize{$^{6}$ Laboratoire d'Ingenierie des Syst\`emes Macromoleculaires, Aix-Marseille Univ, } \\
\normalsize{CNRS, Marseille, France }\\
\normalsize{$^\ast$ Corresponding authors:  jeremie.torres@umontpellier.fr; marco.pettini@cpt.univ-mrs.fr }
}

\date{}

\begin{document}

\baselineskip24pt


\maketitle 


\begin{sciabstract}
Both Classical and Quantum Electrodynamics predict the existence of dipole-dipole long-range electrodynamic intermolecular forces, however these have never been hitherto experimentally observed.  The discovery of completely new and unanticipated forces acting between biomolecules could have considerable impact on our understanding of the dynamics and functioning of the molecular machines at work in living organisms.  
Here, using two independent experiments, based on different physical effects detected by fluorescence correlation spectroscopy and Terahertz spectroscopy, respectively,  we demonstrate experimentally for the first time the activation of resonant electrodynamic intermolecular forces.
 This is an unprecedented experimental \textit{proof of principle} of a  physical phenomenon that,   having been observed for bio-macromolecules and with a long-range of action (up to 1000 ${\mathring{A}}$), could be of importance for biology. In fact, in addition to thermal fluctuations that drive molecular motion randomly, these resonant (and thus selective)  electrodynamic  forces  may  contribute  to  molecular  encounters  in  the crowded cellular space.  
\end{sciabstract}


\section*{Introduction}

Beyond the strong interest for fundamental physics of observing  intermolecular dipole-dipole electrodynamic (ED) forces, this experimental study to detect these forces was highly motivated by their possible role at the molecular level in biology. 
Indeed, from a physicists point of view, living matter offers a wealth of fascinating dynamic phenomena involving biomolecules (proteins and nucleic acids) organized in an intricate and complex network of biochemical interactions providing astonishing efficiency. 
In numbers, a human cell at any given time may contain about 130,000 binary interactions between proteins \cite{Venkatesan:2009}, of which about 34,000 unique human protein-protein interactions are already listed on databases \cite{Bonetta:2010}. The forces hitherto considered in biological contexts are of (quasi-) electrostatic nature (chemical-, and hydrogen-bonds, bare Coulomb, Van-der-Waals London, Hamaker-forces) and are limited to a range of action shorter than 
10 ${\mathring{A}}$ due to Debye screening by small freely moving ions in the intracellular water. 
Such interactions are relevant for stereo-specific, "lock-and-key" and "induced-fit" interactions at short distances but hardly effective to recruit distant molecules. Thus, understanding how 
the right molecule gets to the right place, at the right moment, in the right cascade of events of any biological action is one of the most striking challenges.
Fundamentally, it is assumed that this network of interactions is dominated by random molecular diffusion throughout the cellular spaces in which, sooner or later, a molecule will encounter its cognate partners. However, free diffusion is considerably slowed down in a highly crowded environment \cite{Gori:2016} as in the case of the cell interior. 
Moreover, when diffusion measurements are performed in complex molecular organizations such as those of living cells, most of the biomolecules show anomalous rather than Brownian diffusion \cite{Banks:2005,Golan:2017}. Furthermore, structuring of the cytosol into phase-separated domains \cite{Banani:2017,Berry:2018}, substrate channeling of the metabolons \cite{Sweetlove:2018}, or long-distance interactions in DNA searching \cite{Kulaeva:2012} and organization \cite{Wang:2016} have recently come to the forefront to question the discrepancy between the observed reaction rates in cells with the predictions of a strict random diffusion model \cite{Banani:2017,Weeldon:2016}. 
Within this framework, dipole-dipole electrodynamic interactions are predicted to promote molecular attraction by being selective through resonance and to act over long distances \cite{Preto:2015}. 
These forces can propagate without attenuation in electrolytes of ionic strength comparable to that in the cytoplasm, provided that their oscillation frequency exceeds the Maxwell frequency at few hundreds of MHz \cite{XammarOro:2008}, since at such frequencies the Debye screening is ineffective \cite{Maxwell:1954}.
Therefore, we hypothesize that, in addition to random diffusion, selective long-range attractive electrodynamic interactions increase the encounter rates between $A$ and $B$ cognate partners through a mutual force field described by a potential $U(r)$ where 
$r$ is the intermolecular distance. 
Then, after the Smoluchowski-Debye formula  \cite{Debye:1942,Noyes:1961}, the association rate is given by $k_a^* = 4 \pi R^*(D_A +D_B)$ where $D_{A,B}$ are their diffusion coefficients and $R^*$  is given by
\begin{equation}\label{kdebye}
 R^*=\left[\ {\displaystyle \int \limits_{R_A + R_B}^{\infty} \frac{e^{U(r)/kT}}{r^2} dr \ \ \ }\right]^{-1} .
\end{equation}
where $R_A, R_B$ are the hydrodynamic radii of $A$ and $B$, respectively; $k$ is the Boltzmann constant and $T$ the temperature. Remarkably, for an attractive interaction,  $U(r)<0$, Equation (\ref{kdebye}) implies $k_a^* > k_a = 4 \pi (R_A + R_B)(D_A +D_B)$, that is, an increase of the association rate with respect to the purely diffusive $k_a$.  
However, electrodynamic macromolecular interactions have not previously been considered for several reasons, including: 
\textit{i)} they have never been observed in any experimental context, though theoretically predicted by both Classical \cite{Preto:2015,Frohlich:1977} and Quantum Electrodynamics \cite{QED}; 
\textit{ii)} they require an out-of-equilibrium system \cite{Preto:2015}, which though, of course, found in all living systems, is hard to organize in the laboratory \textit{in vitro}, especially with molecules in metastable states with strongly excited giant dipole vibrations, as required for such interactions; 
\textit{iii)} out-of-equilibrium collective vibrations of macromolecules  are  expected in the $0.1 -1.0$ THz domain, this is experimentally challenging to observe because of the strong absorption of water, and indeed such vibrations have only recently been detected for the first time \cite{Nardecchia:2018} and corroborated by theoretical studies \cite{Kurian:2016,Kurian:2018a}. Interestingly, the coherence of these collective molecular vibrations is theoretically expected to be long-lived \cite{Scully:2019}.

\begin{figure}[h!]
\centering
\includegraphics[scale=0.7,keepaspectratio=true,angle=0]{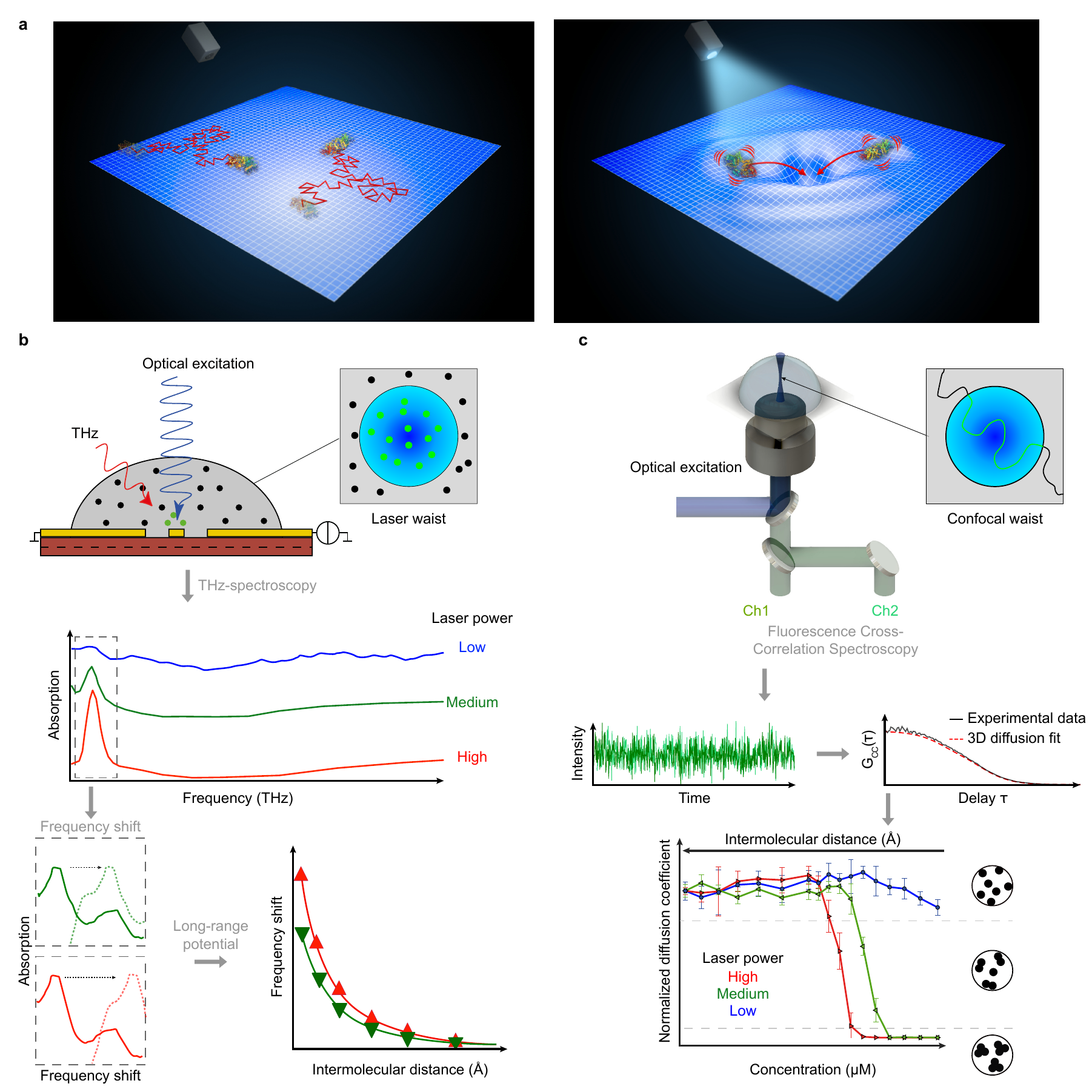}
\caption{\textbf{Long-range electrodynamic interactions - Principle and experimental approaches} (a) At thermal equilibrium, macromolecules show a Brownian diffusive motion in solution (left panel). By switching-on an external energy source, molecules are in an out-of-thermal equilibrium collective vibrational state that can generate ED forces through associated large dipolar resonant oscillations (right panel). (b) In THz spectroscopy, the frequency of the collective vibration varies as a function of the average intermolecular distance (determined by the protein concentration). With respect to the reference frequency at "infinite" dilution, a frequency-shift inversely proportional to the cubic power of the average intermolecular distance is theoretically expected if the proteins interact through ED forces. (c) In FCS, the transit times of the proteins across the volume of observation are measured, hence the protein diffusion coefficients. Under out-of-thermal equilibrium conditions, the theory predicts a phase transition due to ED forces clustering proteins. This should be observed at a given protein concentration by a sudden drop in the diffusion coefficient from its Brownian value. In both (b) and (c), out-of-thermal equilibrium is activated by optical excitation. 
}
\label{prot-prot-interaction}
\end{figure}
Here we demonstrate experimentally the activation of long-range attractive electrodynamic forces between proteins. 
This was made possible through the study of R-Phycoerythrin (R-PE), a protein that can be excited naturally by an external energy supply, a light source. 
Then, by working at  different concentrations (i.e. intermolecular distances) and excitation power of a laser as the light source, we have achieved what follows: \textit{i)} observation of the activation of collective intramolecular oscillation of the proteins, a necessary pre-requisite to activate the physical mechanism  pictorially outlined in Figure \ref{prot-prot-interaction}a; \textit{ii)} observation of a distributed-clustering transition dependent on activation of these collective molecular oscillations, as expected after a thorough preparatory work \cite{Gori:2016,Preto:2012,Nardecchia:2014,Nardecchia:2017}, and \textit{iii)} the consequent expected change of the frequency of the collective oscillation \cite{Olmi:2018}. 
For this, two experimental techniques were used: THz spectroscopy, mainly composed of two setups using either a THz-rectenna or a microwire-probe as sensors, respectively (see Materials and Methods section for details), and fluorescence correlation spectroscopy (FCS) (Figure \ref{prot-prot-interaction}b,c). 
Overall, our experimental work supports a \textit{proof-of-principle} that out-of-equilibrium collective oscillations are capable of activating dipole-dipole electrodynamic intermolecular forces, thus paving the way to explore the potential role of ED intermolecular forces in living matter.

\section*{Results}
This first experimental study has been performed with a natural light-harvesting protein derived from red algae, the R-PE (see Figure S1 in \cite{cite_supp}). 
This protein is an hexamer $(\alpha\beta)_6\gamma$ where the subunit $\alpha$ contains two phycoerythrobilins, the subunit $\beta$ contains three phycoerythrobilins and one phycourobilin and the $\gamma$ subunit contains two or three phycoerythrobilins and one or two phycourobilins. That is, some limited variability in the number of fluorochromes is possible
and $38$ fluorochromes is considered a typical number. All these subunits form one of the brightest fluorescent dye (the entire protein) hitherto found. 
These pigments are highly sensitive to the $\lambda = 488$ nm light used in these experiments. 
\begin{figure}[h!]
\centering
\includegraphics[scale=0.5,keepaspectratio=true,angle=-90]{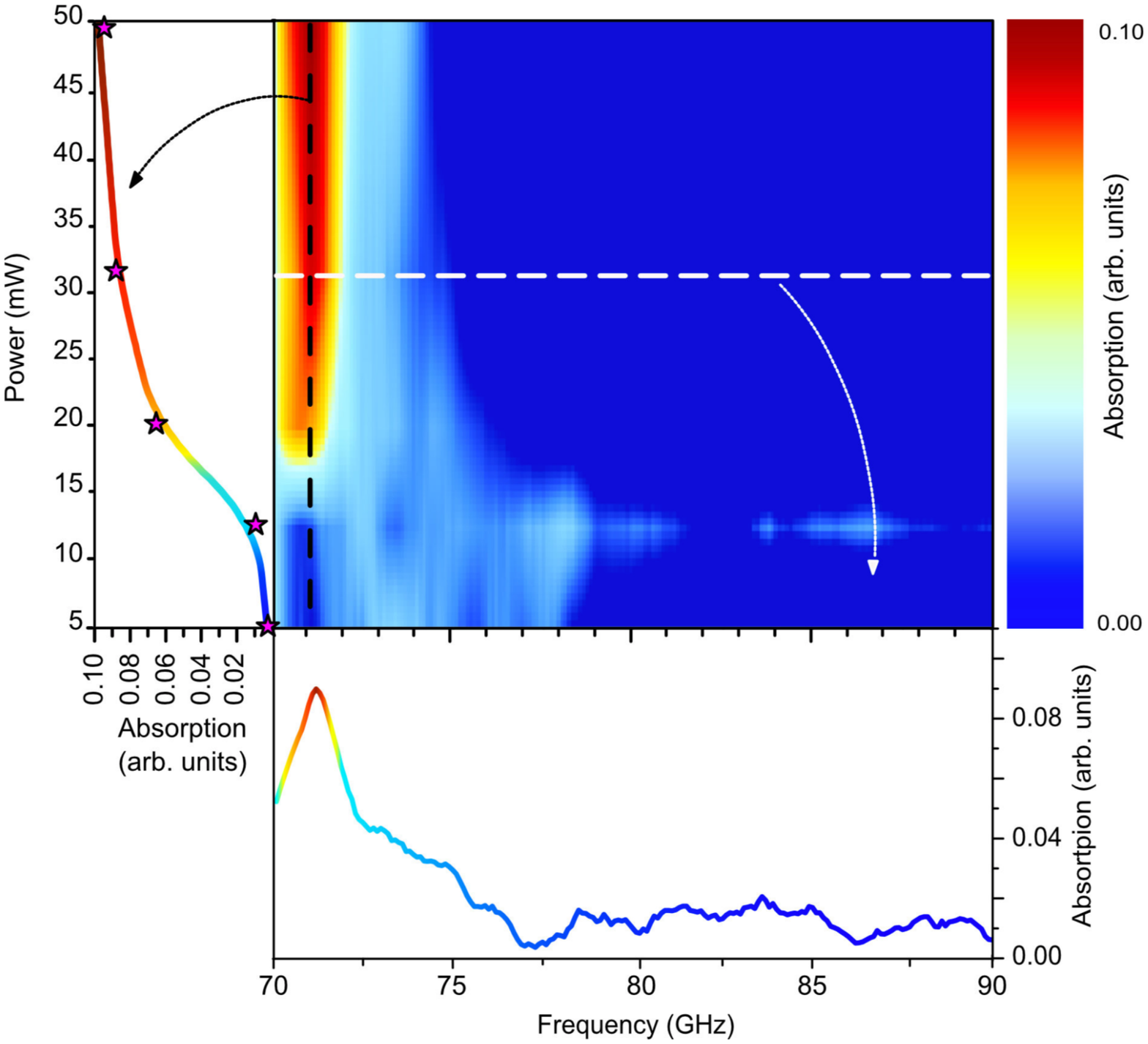}
\caption{\textbf{Collective oscillation of the R-PE measured by THz-Spectroscopy.} Absorption strength of the R-PE (1 mg/mL in 200 mM NaCl) as a function of detection frequency and the laser power activating the molecular collective oscillation at a frequency of 71 GHz (2.4 cm$^{-1}$).
The absorption line profile (bottom) indicated by the horizontal white dashed line was measured at an optical power of 31.5 mW, while the laser power threshold (left) behaviour is indicated by the black dashed line. Stars stand for experimental data 
and the interpolating line is just a guide to the eye. }
\label{RPE_collective}
\end{figure}
\begin{figure}[h!]
\centering
\includegraphics[scale=0.95,keepaspectratio=true,angle=0]{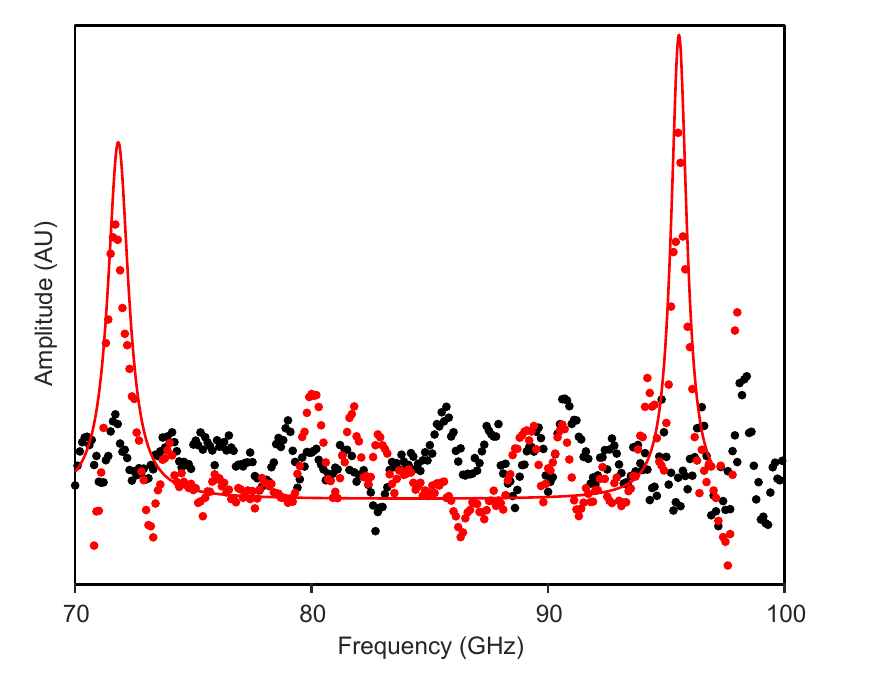}
\caption{{\textbf{ R-PE coherent vibrational states.} Comparison of the R-PE in saline solution (red) and saline solution without R-PE (black). Two collective extension modes of R-PE appear at 71 GHz and 96 GHz. Experimental data (full circles); Lorentz fit (solid line).}} 
\label{RPEcoll-vibr}
\end{figure}


{Thus, under suitable illumination conditions (detail in Materials and Methods) of the R-PE molecules in saline solution, each} molecule can enter a coherent vibrational state (Figure \ref{RPE_collective}) displaying similar phenomena to those reported for Bovine Serum Albumin (BSA) protein  in \cite{Nardecchia:2018}: the existence of a threshold for the energy input rate to activate the collective mode, and a saturation  of the oscillation amplitude at high values of the energy input rate.  
In this saturation regime, two collective oscillation frequencies for R-PE are found (details in Materials and Methods and Supplementary Material). One at 0.071 THz (= $2.4\ {\rm cm}^{-1}$) and another at 0.096 THz (= $3.2\ {\rm cm}^{-1}$),  the latter is only observable under specific saturation conditions of the lower frequency peak.  (see Figure \ref{RPEcoll-vibr}). 
Even if a protein is not an idealized solid, or of strictly well-defined shape, 
considering the best fitting simple geometric shape provided a successful precedent with the BSA. The BSA is referred to as a globular protein and it is certainly not a rigid, elastic, solid sphere, but by applying a formula for the spheroidal modes of an elastic sphere an excellent agreement with the experimental finding was obtained 
\cite{Nardecchia:2018}. Similarly, the geometric shape that best fits the hexamer form of the R-PE is a torus. Even if such a treatment of R-PE and BSA collective vibration modes is somewhat rudimentary, it provides information adequate to what is needed in the present context. In fact, schematizing 
the R-PE protein as a torus-shaped object of $M=240$ kDalton, with a larger midline radius  $R=37.5\ {\mathring{A}}$, a minor radius $r =30\ {\mathring{A}}$ \cite{Liang:1996}, density $\rho$
(the ratio between $M$ and the volume of the torus), and Young elastic modulus $E$, the frequencies of the collective extension modes (those corresponding to oscillating larger radius $R$) are given by \cite{blevins}
\begin{equation}
\nu_n = \frac{(1 + n^2)^{1/2}}{2\pi R}\left(\frac{E}{\rho}\right)^{1/2} .
\end{equation}
Hence the ratio $\nu_1/\nu_0 =\sqrt{2}\simeq 1.41$ which approximates within a $4\%$ the ratio between the observed frequencies, that is $0.096/0.071 = 1.35$. The Young modulus of R-PE is not known in the literature, but by inverting the above formula we obtain $E\simeq 5.3$ GPa which seems reasonable knowing that at 300 K for Myoglobin  $E\simeq 3.5$ GPa, and for BSA $E\simeq 6.75$ GPa \cite{Perticaroli:2013} (both proteins being mostly made of $\alpha$-helices like R-PE). 
{Remarkably, such an empirical approach gives a useful insight on the observed phenomenology as it was successfully done for the BSA \cite{Nardecchia:2018}.}

A theoretically expected signature of attractive electrodynamic forces among
macromolecules vibrating at the same frequency is a phase transition between a dispersed
phase of rapidly diffusing molecules, and a clustered phase of very slowly moving
aggregates of molecules \cite{Nardecchia:2014} (see Supplementary Materials).
The control parameter of this transition is the average intermolecular distance, which
is set by adjusting the molecular concentration.

In the fluorescence correlation spectroscopy (FCS) experiments, the protein  concentrations ${\cal C}$ have been varied in the interval between $0.1\ \mu$M and $10\ \mu$M to make the average intermolecular distance $\langle r\rangle ={\cal C}^{-1/3}$ vary in the interval $550 - 1950\  {\mathring{A}}$. 

The ionic strength of the protein solution is kept at 200 mM by means of suitable concentrations of NaCl in water. This ensures a good shielding of electrostatic interactions.
Moreover, the laser power ($\lambda = 488$ nm) has been varied between  50 and 150 $\mu$W.
Typical fluorescence traces are reported for different laser powers and protein concentrations in \cite{cite_supp} ({Figures S4 and S5)}.

\begin{figure}[p]
\centering
\includegraphics[scale=0.47,keepaspectratio=true]{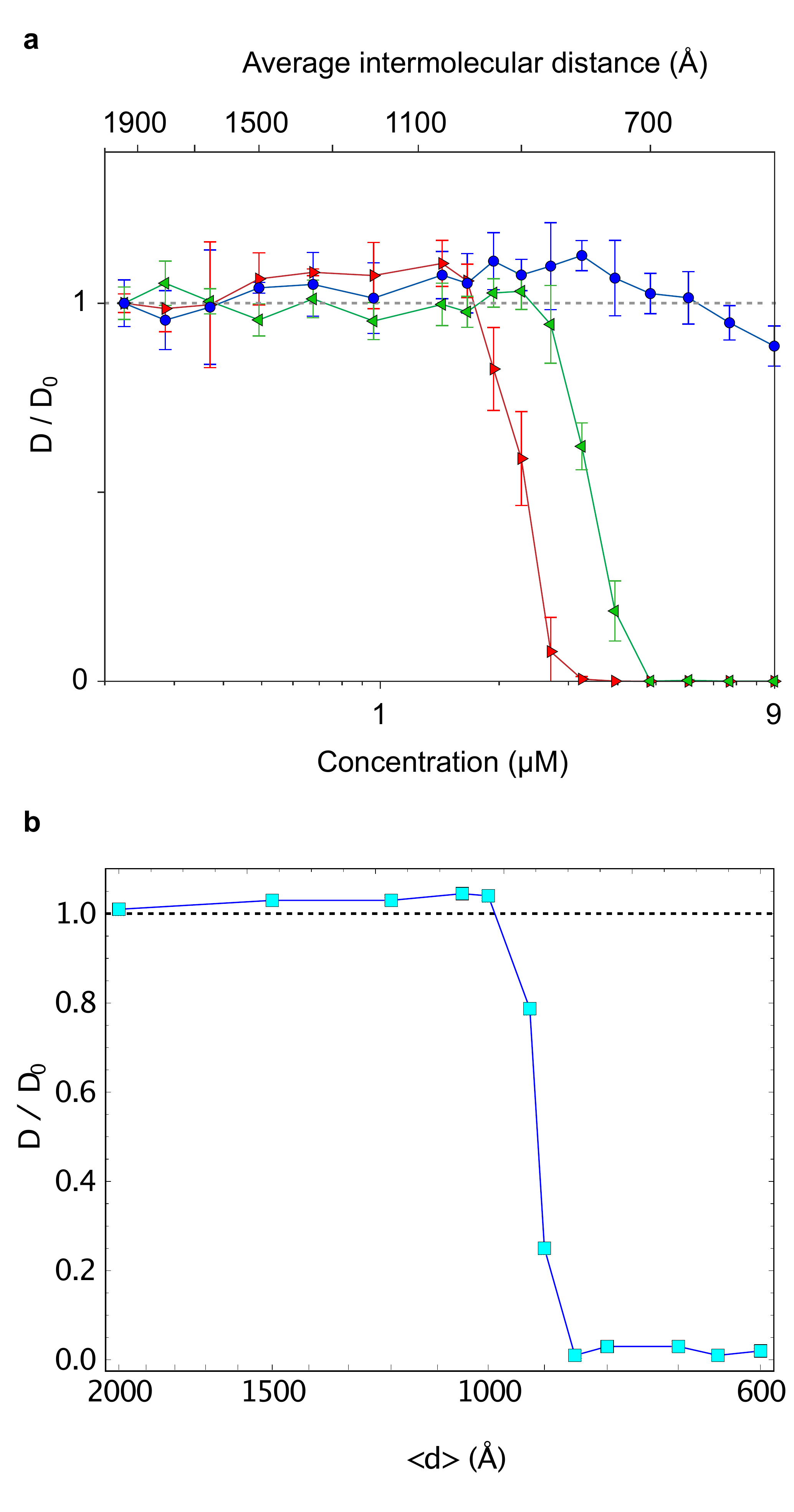}
\caption{\textbf{Effect of protein concentration and laser power illumination on R-PE diffusion: Clustering phase transition.} (a) {Diffusion coefficients normalized to the Brownian $D_0$ values measured for each data series at 0.223 $\mu M$ ($\langle r\rangle \simeq 1950\ {\mathring{A}}$) and recorded at 50 $\mu W$ (blue circles), 100 $\mu W$ (green triangles), and 150 $\mu W$ (red triangles). Each point corresponds to the average of 5 independent experiments. (b)  {Molecular Dynamics computation of self-diffusion coefficient $D/D_0$ (normalized to the Brownian value $D_0$) versus intermolecular average distance $\langle d \rangle$ for a system of particles in a cubic box interacting through long-range electrodynamic forces.}}}
\label{RPE_clusterisation}
\end{figure}

The diffusion times $\tau_D$ determined from the Cross Correlation Functions (CCFs) of fluorescence traces, and the measured waist of the confocal volume where the molecules are both excited and observed, allow  estimation  of the diffusion coefficients $D$. The use of CCFs is motivated in sections Methods and in Additional Data by {Figures S2 and S3} of Ref.~\cite{cite_supp}.    
The values of the diffusion coefficient $D$ are normalized with respect to the Brownian values $D_0$ of each data series recorded at a given laser power (Figure \ref{RPE_clusterisation}). At low laser power the measured value of $D_0$ matches the expected theoretical one $D_0 = k_BT/(6\pi\eta R_H)$, with $\eta$ the viscosity of the solution, and $R_H$ the hydrodynamic radius of the protein.
At laser power of $50\ \mu W$, the observed values of $D$ (blue circles) do not change with the intermolecular distance. The diffusion of the R-PE molecules is Brownian for all the concentrations considered. This means that the energy input rate is either below the threshold value required to excite molecular collective vibrations, or the intermolecular ED forces are very weak because of the small amplitude of the collective vibrations of the R-PE proteins. The mismatch between the laser power values mentioned in Figure \ref{RPE_collective} for THz experiments and those reported above for diffusion experiments is due to the different volumes illuminated (see Methods in \cite{cite_supp}).

At higher laser power, $100\ \mu W$, obviously and consistently with a thermal effect, the measured value of $D_0$ at low concentration is larger than in the preceding case. However it cannot explain, in the interval of inter-molecular distances $700 - 750\ \mathring{A}$ (green triangles), the steep drop of  $D/D_0$ that correlates with the increase of the fluorescence fluctuations and a sudden increase of the diffusion time $\tau_D$.

At the highest laser power considered, $150\ \mu W$, again the steep drop of $D/D_0$ is observed but now in the interval of inter-molecular distances $900 - 950\ {\mathring{A}}$ (red triangles). In both cases, the steep drop of $D/D_0$ correlates with a strong increase of the fluorescence fluctuations and a sudden and huge increase by several orders of magnitude of the diffusion time $\tau_D$.
The strong increase of the fluorescence fluctuations is shown in Figure S6 where, for instance, at an average inter-molecular distance of $600 \mathring{A}$ the variance of the fluorescence fluctuations is more than five orders of magnitude larger than the same quantity measured at an average inter-molecular distance of $1950  \mathring{A}$. Correspondingly, the diffusion times measured through the correlation functions that are reported in the right lower panel of Figure S7 are 4.9 seconds and $4.8\times 10^{-5}$ seconds at an average inter-molecular distance of $600 \mathring{A}$ and of $1950 \mathring{A}$, respectively.
     
The steep drop of $D/D_0$ and the increased amplitude of fluorescence fluctuations are the observable effects of a clustering phase transition stemming from the competition between the electrodynamic intermolecular attractive forces and thermal fluctuations.
This is confirmed theoretically by a semi-analytical model (Figure \ref{RPE_clusterisation}b), by Molecular Dynamics simulations and by Monte Carlo computations (see Supplementary Materials \cite{cite_supp}). 
Molecular dynamics simulations have  been done in order to estimate the effect of long-range electrodynamic interactions on the self-diffusion coefficient $D$ of a system of interacting molecules defined by 
\begin{equation}
D=\lim_{t\rightarrow +\infty}\frac{\langle \|\Delta \mathbf{r}_i (t) \|^2 \rangle_i }{6t}
\end{equation}
where $\Delta \mathbf{r}_i(t)$ is the displacement at time $t$ of the $i$-th molecule with respect to its initial position, and $\langle \cdot \rangle_i$ is the average over all the particles in the system.
This is the physical quantity measured by means of FCS experiments thus allowing a direct comparison between the outcomes of  numerical simulations and the outcomes of lab experiments.

The dynamics is described by the Langevin equations in the overdamped limit (without inertial terms):
\begin{equation}\label{stocEqs}
\frac{\mathrm{d}\mathbf{r}_i}{\mathrm{d}t}=-\frac{1}{\gamma}\nabla_{\mathbf{r}_i} \sum_{j\neq i} U(\|\mathbf{r}_i-\mathbf{r}_j\|)+\sqrt{\frac{2 k_B T}{\gamma}}\boldsymbol{\xi}_i(t) \qquad \forall i=1,..,N
\end{equation}
where $\gamma$ is the viscous friction constant, $U$ is the ED interaction potential, $T$ is the temperature of the solution and
$\boldsymbol{\xi}_{i}(s)$ is a noise term accounting for the random collisions of water molecules, s.t. 
\begin{equation}
\langle \xi_{A,i}(t) \rangle_t=0 \quad \langle \xi_{A,i}(t) \xi_{B,j}(t') \rangle_t=\delta(t-t')\delta_{AB}\delta_{i,j} \qquad \forall i,j=1,...,N \quad \forall A,B=1,...,3\,\,.
\nonumber
\end{equation} 
The parameters entering the model are chosen to reproduce the experimental conditions (details are in Supplementary Material) and the sudden drop of the diffusion coefficient is found in correspondence with the formation of clusters at an average intermolecular distance in agreement with the experimental finding at 150 $\mu W$ in Figure \ref{RPE_clusterisation}a.
\begin{figure}[h!]
    \centering    \includegraphics[scale=0.45]{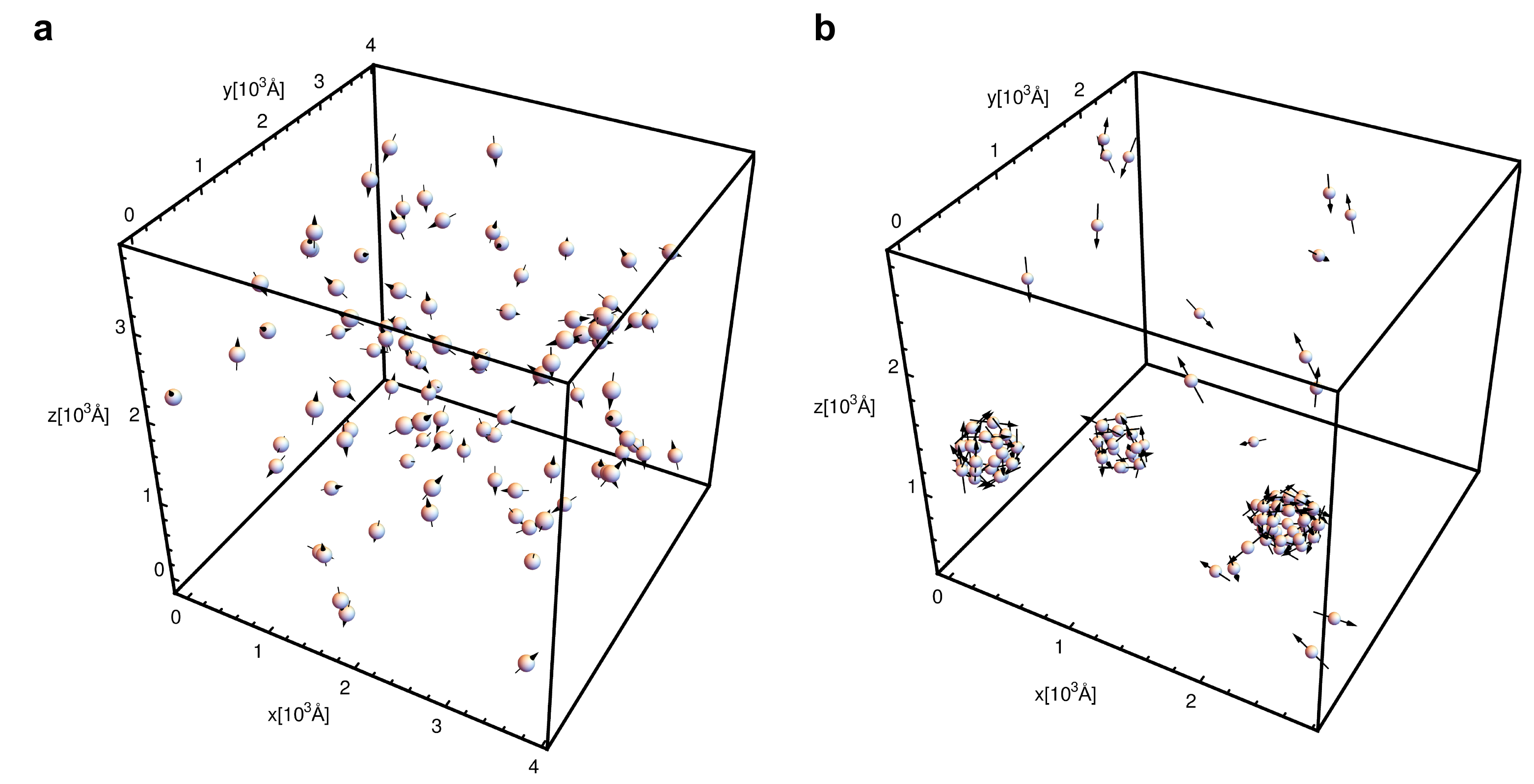}
    \caption{{\textbf{Snapshots of numerical Monte Carlo computations}. Clustering transition induced by long-range electrodynamic inter-particle forces observed by changing the initial value of $\langle r\rangle$. With $\langle r\rangle_{initial} \simeq 1000\  {\mathring{A}}$ the system remains in the dispersed phase (left box) where $D/D_0=1$ (Brownian diffusion); with $\langle r\rangle_{initial} \simeq 950\ {\mathring{A}}$ the system switches to the clustered phase where $D/D_0$ is very small (right box).}}
    \label{fig:my_label}
\end{figure}
In {Figure \ref{fig:my_label}} two snapshots are shown of numerical Monte Carlo computations performed by considering long-range electrodynamic interparticle forces. Thus, the appearance of clusters predicted theoretically has been experimentally observed through the sudden enhancement of fluorescence fluctuation amplitudes and long transit times across the confocal volume of the FCS setup.
Clusters of R-PE were visually evidenced  by fluorescence microscopy in Figure \ref{screenshots-XY} and Video in online Supplementary Materials.  

Increasing the laser power, the higher the power the stronger the collective oscillation of R-PE molecules and the larger the associated oscillating dipole moment. Hence the displacement of the clustering transition to a greater average intermolecular distance for the stronger molecular oscillations. 
Consistently, by suddenly lowering the power of the laser light supply in the clustered phase a rapid disaggregation of the clusters is observed, as shown   
by the sequences of frames reported in Figure \ref{screenshots-XY}. Note that the frame on the left top of this  figure shows that, in the same conditions of temperature and particle concentration,  when the energy input rate is below the threshold value to activate intramolecular collective vibrations the clustering transition disappears. 

All these results are a clearcut proof of the activation of electrodynamic intermolecular attractive forces due to the collective vibrations of the R-PE molecules.
\begin{figure}[h!]
\centering
\includegraphics[scale=1.75,keepaspectratio=true]{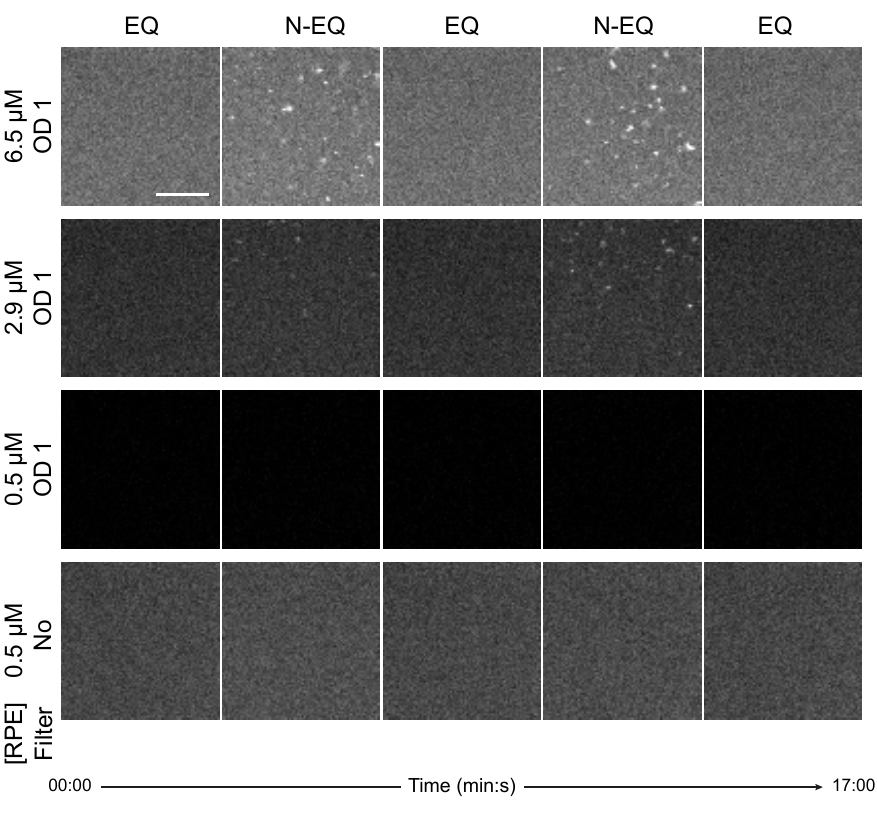}
\caption{\textbf{Screenshots saved out of online videos of a solution of R-PE under laser light illumination}. Optical power alternatively varying in time at 50 $\mu$W  and 150 $\mu$W. EQ stands for proteins at equilibrium (switch-off of collective oscillations) and N-EQ for out of equilibrium (switch-on of collective oscillations). Snapshots have been taken for a concentration of 0.5 $\mu$M (average molecular distance of $\langle d\rangle = 1500\ \mathring{A}$) with and without optical density filter showing the absence of cluster formation at low concentration. At concentrations of 2.9 $\mu$M ($\langle d\rangle  = 800\ \mathring{A}$) and 6.5 $\mu$M ($\langle d\rangle  = 630\ \mathring{A}$) the formation of protein clusters and their reversibility are well evident. 
The increase of the background brightness in the upper sequence of frames is due to the combined effects of the increase of protein concentration and of the laser power. The scale bar corresponds to 10$\mu$m. The units of the time scale are expressed in min:sec. }
\label{screenshots-XY}
\end{figure}
An independent confirmation of the activation of electrodynamic intermolecular attractive forces is obtained by measuring the shift of the collective vibrational frequency as a function of the  concentration of molecules. Once again the proteins were observed in aqueous solution with the
addition of 200 mM of NaCl to screen electrostatic interactions.
In fact, theory predicts that electrodynamic dipole-dipole interaction between two molecules results in a shift $\Delta\nu$ of their vibration frequency from the unperturbed frequency $\nu_0$, shift proportional to $1/r^3$ with $r$ the intermolecular distance \cite{Preto:2015}. 
Remarkably, this law is preserved also for a large number of interacting molecules, that is, the frequency shift is proportional to $1/\langle r\rangle^3$ where $\langle r\rangle$ is the average intermolecular distance given by $\langle r\rangle ={\cal C}^{-1/3}$  (see Supplementary Materials for the theoretical explanation \cite{cite_supp}). The unperturbed vibration frequency $\nu_0$ is operationally measured at very low molecular concentration. The shifts $\Delta\nu$ are then measured with respect to this value of 
$\nu_0$.  
\begin{figure}[h!]
\centering
\includegraphics[scale=0.77,keepaspectratio=true]{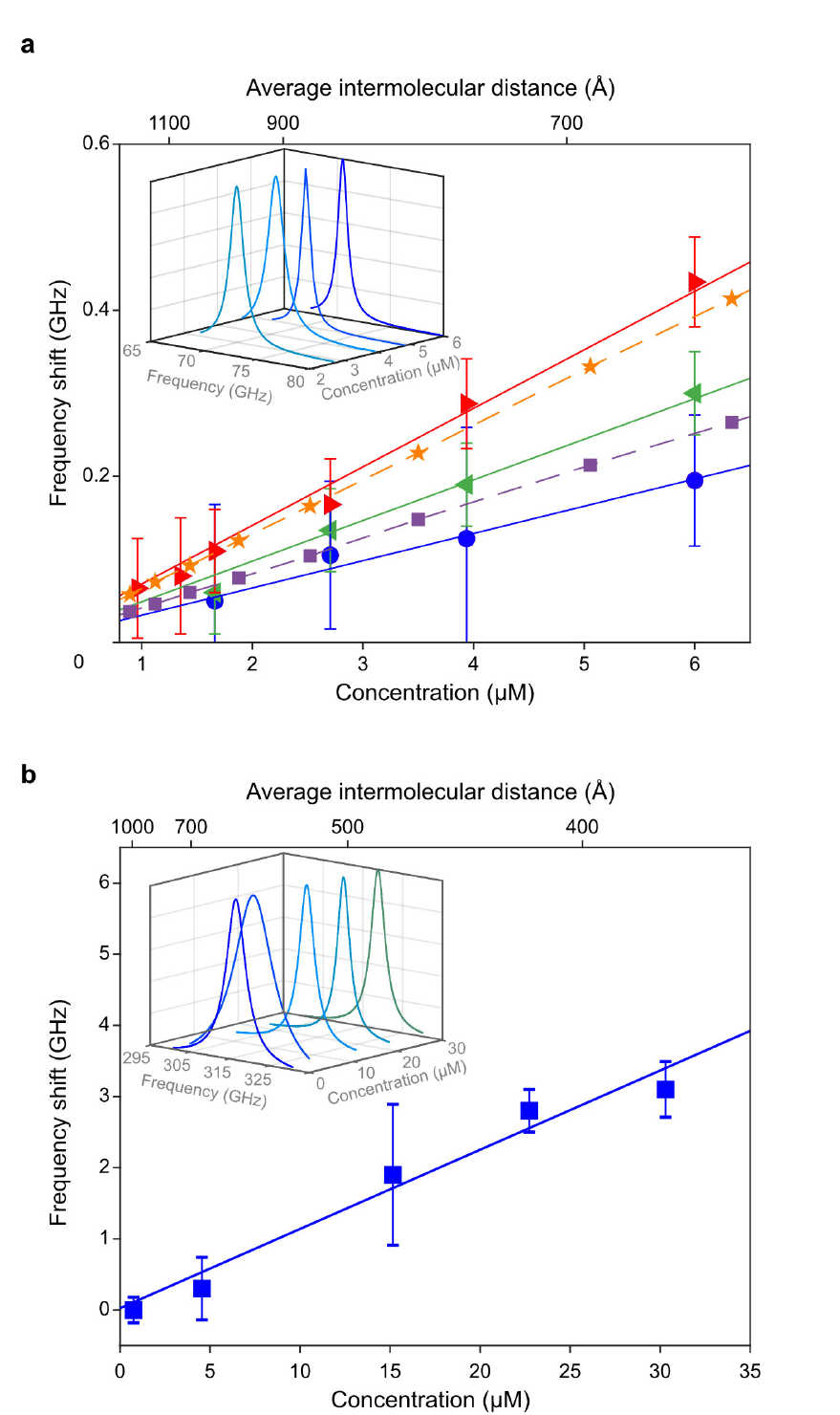} 
\caption{\textbf{Frequency shifts of the intramolecular collective vibrations of R-PE and BSA at different concentrations.} {Measurements were performed at room temperature in aqueous solution with 200 mM of NaCl. Panel (a) refers to R-PE. The shift is relative to the reference frequency measured at the lowest  protein concentration. Measurements have been performed  at different powers of the laser: 31.5 mW (blue circles), 39.5 mW (green triangles), 50 mW (red triangles). Purple squares and orange stars refer to theoretical outcomes worked out with different values of molecular dipole moments (see Supplementary Materials). Panel (b) refers to BSA at a laser power of 40 mW.}  Insets: The frequency shifts are measured through a Lorentz fitting of the experimental resonances. The different colors in the insets are just a visual help.  }
\label{f-shift}
\end{figure}
The collective oscillation frequency of the R-PE at 0.071 THz (i.e. $2.4\ {\rm cm}^{-1}$) allowed to use as
THz detector an electronic nano-device, namely a bow-tie antenna on the 2D electron gas layer
of a FET transistor called a rectenna \cite{Schuster:2011} 
(see Materials and Methods section). 
The measured frequency shift is found to follow a linear
dependence on the protein concentration, that is, the experimental outcomes are in excellent
agreement with the theoretical predictions (Figure \ref{f-shift}a).
The steepening of the fitted lines in Figure \ref{f-shift}a  as laser
power is increased is also in agreement with expectations.
This is because the increasing oscillation
amplitude (entailing a larger oscillating dipole moment), results
in stronger intermolecular electrodynamic interaction and hence a larger frequency shift (for
theoretical details see Supplementary Materials \cite{cite_supp}). The case of R-PE is especially important
because two independent experimental approaches lead to the same physical conclusion: the
activation of intermolecular electrodynamic forces acting at a long distance. Thus the possibility
of artefacts or misinterpretations is ruled out. Importantly, these forces are not peculiar of the R-PE since they are also observed by making measurements with another protein, BSA.
BSA molecules were labeled with an average number of 5 fluorochromes AF488 5-TFP, covalently bonded to the Lysine residues, excitable by the light
emitted by a diode laser at 488 nm. The reason for considering the BSA is twofold: first, the
labeled BSA molecules, when continuously supplied by laser light at $\lambda = 488$ nm, are driven
in a collective vibration mode at 0.314 THz  (i.e. $10\ {\rm cm}^{-1}$) \cite{Nardecchia:2018}; second, R-PE and BSA both have a structure composed largely of alpha-helices (Figure S1 in \cite{cite_supp}),
a property that we can suppose being facilitating the activation of collective vibrational modes. In fact, alpha-helices behave as springs undergoing the so-called accordion modes, therefore a protein mainly composed of alpha-helices is reasonably supposed to be highly flexible and its components can undergo synchronization phenomena possibly giving rise to a collective oscillation.
The collective oscillation frequency of the BSA at 0.314 THz required the use of a near-field microwire-based probe coupled with a wave-guide (see Figure \ref{THz_spectro}b of Materials and Methods section).
Again, the measured frequency shift is found to follow a linear dependence on the protein concentration as expected (Figure \ref{f-shift}b). This provides a remarkable confirmation of the activation of long distance electrodynamic interactions also for BSA.
 
It is worth pointing out that the two THz experimental setups have been operated at a different spatial density of the laser light, with respect to diffusion experiments, to prevent a clustering transition during the frequency shift measurements.  In fact, by using a set of optical density filters to lower the spatial density of the laser power, the ED forces can be weakened to the point of being unable to entail the clustering transition by overcoming thermal noise, and the protein molecules keep diffusing with a Brownian motion (see Materials and Methods section). 

\section*{Discussion}
In the  work  presented  here,  we  have experimentally shown,  for  the  first  time,  that  the  excitation  of  out-of-equilibrium  collective  oscillations  is  capable  of  driving molecular  association  through  the activation of electrodynamic intermolecular forces.  
R-PE provides convincing evidence for these forces using a pair of complementary experimental approaches. 
Thus, R-PE represents the \textit{``Rosetta stone"} allowing us to identify attractive intermolecular forces through two completely different physical effects. 

Long-range electrodynamic interactions have also been activated using a second protein, fluorescent dye-labeled BSA.
These results confirm the generality of this phenomenon thus paving the way for the experimental searching of ED forces in the more complex environment encountered in cell biology.

A central feature of the ED forces activated by collective molecular vibrations is their long-range property, that is, they stem from an 
interaction potential decreasing as the third inverse power of the intermolecular distance. This  generates first-order phase transitions with the formation of molecular condensates. A thorough theoretical analysis of this transitional behavior, and of the frequency shift of the collective vibrations of the proteins, caused by their interaction through ED forces, can be found in Supplementary Materials.

By being dynamic and reversible, ED forces can be instrumental in structuring the mesoscale molecular organization through the formation of biomolecular condensates and thus provide a rationale to explain the speed of many cellular processes. While we have evidence of the association of biological macromolecules driven by ED forces \textit{in vitro}, the conditions are far from the ones found  \textit{in cellulo}. Here, the energy sources necessary to maintain the molecules out-of-equilibrium need to be characterized. Among the potential candidates, there are:  adenosine triphosphate (ATP) as a universal biological fuel, ionic currents such as those used to drive ATP synthesis, ionic collisions, photons produced by mitochondria, or external light
\cite{VanWijk,Celardo_2019}. The notion of selectivity of ED forces will also need to be explored in a crowded environment composed of a vast diversity of molecular components. More specifically, it will be important to consider the possible role of ED forces between different molecular species (ligand-receptor, DNA-protein) co-resonating at one or more collective oscillation frequencies. 
Interestingly, recent numerical studies have shown that correlations of electronic
density fluctuations can extends up to \SI{30}{\angstrom} in protein-solvent 
systems \cite{stoer:2019}. These correlations, responsible for long-range many-body van der Waals interactions, by structuring hydration layers, could support and give a contribution also to ED intermolecular interactions and comunications among biomolecules.

Moreover, our findings could be of general interest for methodological applications requiring a tight control in space and time of molecular interactions in living cells to decipher cellular functions. As an example, in the developing field of optogenetics where molecular interactions are switched on/off by engineered proteins controlled by light \cite{Tischer:2014}.

Further, ED forces could also be exploited in the domain of drug design. It has been hypothesized that combining steric complementarity with ``long-range complementarity" could improve therapeutic efficiency. Thus, ED forces could allow recognition at a distance and drive an efficient mutual approach between a drug and its therapeutic target before their final chemical binding  \cite{Veljkovic:2011}.

Finally, throughout evolution, life has exploited all the available physical laws, processes, and phenomena. It is thus exciting to speculate how the new phenomenon reported here for the first time has been used to gain adaptive advantage and possibly overcome some of the adverse effects of molecular crowding in cells.

\section*{\large Materials and Methods}

\subsection*{Theoretical rationale}
A crucial condition to activate long-range electrodynamic forces between biomolecules is to put them out-of-thermal-equilibrium, in a coherent vibrational state of all - or a large fraction - of their atoms. 
In fact, at thermal equilibrium a macromolecule "flickers", that is, it undergoes incoherent and random deformations of all its subunits resulting in small noisy variations of its static dipole moment. At equilibrium, energy equipartition among the normal modes of vibration holds, thus the amplitude of collective modes is very small. Collective modes have been detected at equilibrium already a long time ago for proteins \cite{Painter:1982} and polynucleotides \cite{Painter:1981} with Raman and far infrared spectroscopy. 
But by keeping a macromolecule out of thermal equilibrium by means of an external energy supply a phonon condensation phenomenon can be activated \cite{Nardecchia:2018}, consisting of the channelling of the supplied energy into a low frequency  coherent mode of vibration of all the atoms (or of a significant fraction of them) of the same macromolecule. So excited, the macromolecule may have a very large oscillating dipole moment. In turn, in such a collective oscillation state, a biomolecule could activate a long-range attractive electrodynamic interaction with other molecules if they possess a proper frequency close to that of the coherent mode. This scenario has been proposed a long time ago 
\cite{Frohlich:1968,Frohlich:1970,Frohlich:1972,Frohlich:1977} but has been marginalized for several reasons, including its quantum mechanical formulation and because it has never been given experimental evidence. Reformulated in a classical framework \cite{Preto:2015,Nardecchia:2018}, the above sketched scenario has been given new credit after the recent experimental proof of the possibility of activating a stationary oscillation of a macromolecule out of thermal equilibrium  \cite{Nardecchia:2018}. This is a necessary pre-condition to activate long-range electrodynamic interactions \cite{Preto:2015}.
It is worth mentioning that in a recent work it has been convincingly maintained that a full understanding of Fr\"ohlich condensation can come solely from a quantum theory \cite{Scully:2019}.

\subsection*{Theoretical design of the experiments}
{Classical methods for measuring the forces between molecules and characterizing them, such as AFM or optical tweezers, even if they give very interesting results on molecular interactions, are not the  most obvious and easy to implement in the specific cases of out-of-equilibrium oscillating molecules. Indeed, any technique requiring to bind a macromolecule, like a protein, to nano-objects such as nanobeads, nanowires or surfaces would have an unpredictable effect on the molecular collective oscillations. To go further we have thus designed other specific methods to detect long-range electrodynamic intermolecular forces.
A first method conceived is based on the expected change} in the diffusion properties of the proteins at different concentrations, in the presence of both suitably activated electrodynamic forces and stochastic forces representing the random hits of water molecules on the proteins.
The feasibility study of this approach has been carried out via theoretical computations resorting to Molecular Dynamics simulations.  
Experimentally, the FCS/FCCS methods that are based on recording the fluctuations of fluorescence signal in a small volume of observation (i.e., a confocal volume in FCS microscopy) have the sensitivity to infer information about the diffusion time through this volume \cite{Haustein:2007}.
The autocorrelation (cross-correlation) function of the detected fluorescence signal is fitted by means of an analytic formula
derived by a theoretical model of diffusion. Hence the diffusion time $\tau_D$, that corresponds to the average time molecules stay within the volume of observation, is obtained and the diffusion coefficient $D$ is simply given by
 $D = \omega^2_{xy}/4\tau_D$, where $\omega^2_{xy}$ is the lateral waist of the excitation beam focused through the objective of the microscope. This allows us to assess whether the transit time of fluorescent molecules follows a Brownian diffusion law (i.e., resulting from stochastic forces generated by collisions with water molecules) or, instead, a law that possibly combines both electrodynamic forces and the stochastic forces just mentioned. A three steps feasibility study has validated this experimental strategy to detect these electrodynamic intermolecular forces 
  \cite{Preto:2012,Nardecchia:2014,Nardecchia:2017}.
In particular after the results reported in Ref.\cite{Nardecchia:2014}, the free Brownian diffusion was expected to be dramatically reduced by the formation of large molecular clusters when the concentration of the solvated molecules
exceeds a critical value. These preliminary results have 
suggested to experimentally look for this hallmark of the activation of electrodynamic interactions. However, a more advanced and thorough investigation of this clustering transition was in order, thus it has been done in the present work (see Section 2.3 of the theoretical part of this Supplementary Material). By resorting to a semi-analytical model, to Molecular Dynamics simulations, and to Monte Carlo computations it is shown that the activation of electrodynamic  forces entails the existence of a first-order clustering phase transition. This transition between a "dispersed" phase and a "clustered" phase occurs at a critical value of the concentration of actively oscillating biomolecules out-of-thermal-equilibrium, therefore at a critical value of the average intermolecular distance.
An independent and complementary experimental possibility of detecting the activation of intermolecular
electrodynamic  forces has been suggested in Ref.\cite{Preto:2015}.
In that case, the frequency of the collective oscillation mode of an isolated macromolecule undergoes a shift when placed close to another macromolecule oscillating at the same frequency. This shift is inversely proportional to the cube of the intermolecular distance. Remarkably, this remains true for a large number of molecules in solution (see Section 3 of the theoretical part of this Supplementary Material) in which case the frequency shift is found again to be inversely proportional to the cube of the \textit{average} intermolecular distance. If this represents an experimental crosscheck of the diffusion-based approach, challenge comes \textit{a-priori}  from the difficulty in implementing THz spectroscopic measurement of proteins in aqueous solutions due to the strong absorption of water in the THz and sub-THz frequency domains. But this difficulty is considerably reduced when measurements are performed under out-of-equilibrium conditions. In fact, this is distinct from standard spectroscopic measurements, where the incident radiation simultaneously excites the atoms or molecules under study and probes their absorption spectrum by sweeping a range of frequencies. With THz spectroscopy of out-of-equilibrium biomolecules in solution, the analysis is carried out on already active objects (whose vibrations are induced by an internal cascade or interconversion of light), and weak THz radiation is only used to read and detect active vibrations of the biomolecules. 
Moreover, the absorption of THz radiation by the actively oscillating molecules is much stronger than the absorption of THz radiation by water, thus making molecular absorption features well detectable despite the  presence of water.

\subsection*{Two experiments operating in different conditions}
In the FCCS setup, the laser beam is focused in a small volume of about one femtoliter where a high energy density is attained. This is necessary to excite the collective oscillation of the R-PE molecules during their quick transit through the confocal volume of transversal diameter of 916 nm. The R-PE molecules efficiently harvest light by means of their 38 fluorochromes within each protein molecule.
On the contrary, in the THz spectroscopy setup, a much lower energy density was attained because the laser beam illuminated a large drop of solution of 35$\ \mu$L.
This notwithstanding,  the excitation of collective oscillations of both the BSA molecules and of the R-PE molecules, respectively,  was attained - without making them cluster - by means of long exposure times. The activation of collective oscillations of the molecules in solution preventing at the same time their clustering (which simply means that the electrodynamic forces were unable to overcome the thermal forces) was a necessary condition, in fact, a clustered fraction of the molecules would have completely altered the relationship between frequency shift and concentration.  This operating condition was obtained by adjusting the laser power so as to lower the threshold value of the average intermolecular distance of the clustering transition below $500\ {\mathring{A}}$ for the R-PE. Moreover, the same laser power density was unable to induce the clustering phase transition of the BSA molecules because the power density was not enough  to excite sufficiently strong collective oscillations. This is the reason why the BSA turned out not apt for the experiments based on self-diffusion. In fact, the BSA is not naturally sensitive to light, and, to get excited, its collective oscillation mode requires a very long time of light-pumping, via the artificially attached fluorochromes, 
of about 10 minutes \cite{Nardecchia:2018} to make the corresponding absorption feature sharp enough. Long excitation times are necessary because, in our experimental conditions and for BSA, the rates of energy dissipation and energy input are almost equal. A long time is therefore needed to accumulate enough energy in each protein to make intra-molecular nonlinear interaction terms strong enough to activate the condensation phenomenon. For R-PE, which naturally absorbs light and contains a total of 38 fluorochromes, the process is much faster, and a time of less than three seconds is necessary to observe the formation of clusters in the FCS experiments, whereas in THz experiments about one minute  (at 50 mW)  is needed to begin to detect the spectral features corresponding to collective molecular vibrations and about 4 minutes are necessary to observe a sharp absorption peak  at 71 GHz and about 15 minutes to excite the second peak at 96 GHz. The physical reason for this is that the phonon condensation phenomenon - that activates the collective vibration of each protein molecule - consists in channelling all the injected energy into the fundamental vibrational mode. Only when the  saturation regime is attained some energy becomes available also for higher frequency collective vibrational modes. As above discussed, the phonon condensation is activated after a suitable exposure time of energy injection, that is, when enough energy is  accumulated in each molecule. The experimental fact that a much longer exposure time is required to observe the higher frequency peak is thus consistent.  In both cases, the mentioned values of time pumping is highly reproducible. 

Long illumination times were realised in the THz setup but this was not possible in the FCS/FCCS setups 
where the transit times throughout the confocal volume was exceedingly short to activate the collective oscillations of the BSA molecules in spite of the high light power density in the confocal volume. In conclusion, let us remark that the electrodynamic forces can be active also in the disperse phase when the proteins undergo Brownian diffusion. In other words, even if these forces are not strong enough to make the clustering transition by winning against thermal disorder, they can nevertheless induce the frequency shifts reported in Figure 4.


\subsection*{\bf Biochemical samples}
R-PE was purchased from Merck (52412). BSA conjugated to Alexa Fluor$^{TM}$ 488  was purchased from Thermo Fischer Scientific (A13100). Samples were prepared by sequential dilutions of proteins in 200 mM NaCl. Their final concentrations were controlled by absorbance measurements (Nanodrop One Spectrophotometer, Thermo Fischer Scientific) at 566 nm for R-PE ($\varepsilon = 1\ 960\ 000\ {\rm M}^{-1}{\rm cm}^{-1}$) and at 280 nm for BSA ($\varepsilon = 43 824\ {\rm M}^{-1}{\rm cm}^{-1}$ ) with a correction factor of 0.11 to account for AF488 at 280 nm.

\subsection*{Fluorescence Correlation Spectroscopy experiments}

 \textit{\textbf{Setup}}. All fluorescence correlation spectroscopy measurements were performed on a confocal microscope (ALBA FCSTM, from ISS Inc., Champaign, USA) with a picosecond/CW 488 nm diode laser (BDL-488-SMN, Becker and Hickl, Germany) used at 80 MHz and focused through a water immersion objective (CFI Apo Lambda S 40X/1.25 WI, Nikon). The fluorescence collected by the same objective is split into two paths by a 50/50 beam splitter (Chroma 21000) and filtered by 525/40 nm band pass (Semrock FF02-525/40) before been detected by avalanche photodiodes (SPCM AQRH series, Perkin Elmer/Excelitas). Signals are recorded by a multitau hardware correlator (FLEX-02-12D, Correlator.com, Bridgewater, NJ).
 
\noindent \textit{\textbf{Data acquisition}}.  Before each experiment, the laser power is adjusted at the back-aperture objective and, the lateral waist value $\omega$ calculated knowing the diffusion coefficient of AF488 [$D_{AF488}(20^{\circ}C) = 409\ \mu$m$^2/$s derived from \cite{Petrek:2008}] in aqueous solution at $20^o$ C and according to the equation $\omega^2 = 4 D\ \tau_D$.
FFS measurements were performed on sample in 8-well LAB-TEK chambers (Thermo Scientific Nunc) at given concentration of proteins diluted in 200 mM NaCl. Five independent experiments were conducted for each condition (protein concentration and laser power) with each individual measure corresponding to a series of 10 measurements lasting 60 seconds. 
Different strategies have been previously applied to perform FCS measurements at micromolar concentrations \cite{Laurence:2014, Khatua:2015}. Here, to overcome this concentration limitation, which is mainly due to the detector limits, the collected signal was attenuated by optical density (OD) filters on the path just after the emission filters. The addition of OD1 or OD1.3 filters with transmission values of $10\%$ and $5\%$, respectively, allows keeping records with satisfactory amplitude fluctuations and count rates per molecule for reliable analyses. The effectiveness of this experimental approach has been validated with high concentrations of aqueous solutions of the Atto488 dye (Additional Data Figures S2 and S3). In addition, the use of the FCCS data acquisition modality avoids post-pulse artefacts due to spurious photon detection \cite{Zhao:2003}.

\noindent \textit{\textbf{Data analysis}}.  As mentioned above, all data were collected in a FCCS mode for proper experimental analyses. Consequently, we analyze the fluctuations by a cross-correlation function (CCF) defined as 
\begin{equation}
G({\tau})=\frac{\langle {\delta}F_{1}(t){\delta}F_{2}(t+{\tau})\rangle}{{\langle}F_{1}{\rangle}{\langle}F_{2}{\rangle} }\, ,
\end{equation}
with $\delta F_1(t)$ and $\delta F_2(t)$ the signal collected by detector 1 and 2, respectively. The averages are performed over time.
We kept  simplest the mathematical model fitting the CCF by using the standard analytic form of the function $G({\tau})$ for a one-component system in a 3D environment which reads as \cite{Magde:1974}
\begin{equation}
G({\tau})= \frac{1}{N} \left(\frac{1}{(1+{\tau}/{\tau_D})\sqrt{1+s^2{\tau}/{\tau_D}}}\right),
\label{diffusion_equation}
\end{equation}
where $N$ is the average number of molecules, $\tau_D$ the average diffusion time of the tracer through the confocal volume, and $s$ the structure parameter of the confocal volume, that is, its axial to lateral waist ratio $\omega_{xy}/\omega_z$. 
Still, the fitting accuracy of CCFs with Eq.(\ref{diffusion_equation}) is no longer very good with high laser power and at high concentrations (Additional Data Figure S7) because of CCF distortion due to very long transit times of protein clusters. Therefore, the CCFs that are no longer accurately fitted by the standard function $G({\tau})$ are  preliminarily smoothed by applying a Savitzky-Golay filter using a polynomial of third degree (red curves on Additional Data Figure S7). This allows to get reasonable estimates of $\tau_D$ as the lag time corresponding to the half-height of the smoothed CCF. Note that when the R-PE concentration exceeds even slightly a clustering critical value, these estimates are fully acceptable and adequate for our purposes. In fact, in this case $\tau_D$ suddenly increases of several orders of magnitude with respect to its Brownian value, thus making the details of its estimation irrelevant. The variation of $\tau_D$ is also so large that the effects of the high level of noise at short lag-times on its  estimate do not have any meaningful consequence on the clear evidence of the clustering phase transition.

\subsection*{Confocal video-microscopy} 
The confocal video-microscopy images were performed on a confocal microscope (ALBA FCSTM, from ISS Inc., Champaign, USA) with a picosecond/CW 488 nm diode laser (BDL-488- SMN, Becker and Hickl, Germany) used at 80 MHz and focused through a water immersion objective (CFI Apo Lambda S 40X/1.25 WI, Nikon). The fluorescence collected by the same objective is filtered by  a 525/40 nm band pass (Semrock FF02-525/40) before being detected by avalanche photodiodes (SPCM AQRH series, Perkin Elmer/Excelitas). 
Each $128  \times 128$ pixel confocal image correspond to a $30  \times 30$ $\mu$m field of view and recorded at a pixel dwell time of 0.1 ms at a rate of 2.8 s/frame. 
\subsubsection*{Captions for Movies  S1 - S4}

\noindent\textbf{Mov1-seq1-v1} :
R-PE protein concentration 0.5 micro Moles (absence of protein clusters). No optical filter. Each 128 x 128 pixel confocal image corresponds to a 30 x 30 $\mu$m field of view and recorded at a pixel dwell time of 0.1 ms at a rate of 2.8 s/frame. Each stack of images was made of 375 images where the laser power has been adjusted every 75 images to 50, 150, 50, 150 and then 50 $\mu$W. When indicated, an OD1 filter (Thorlabs ND-10B) has been added before the detector. Then, the video made from these images were compressed at a rate of 7 frame/s. Thus, each second of video corresponds to about 21 seconds of measurements.

\noindent\textbf{Mov2-seq2-v1} :
R-PE protein concentration 0.5 micro Moles (absence of protein clusters).  Each 128 x 128 pixel confocal image correspond to a 30 x 30 $\mu$m  field of view and recorded at a pixel dwell time of 0.1 ms at a rate of 2.8 s/frame. Each stack of images was made of 375 images where the laser power has been adjusted every 75 images to 50, 150, 50, 150 and then 50 $\mu$W. An OD1 filter (Thorlabs ND-10B) has been added before the detector. Then, the video made from these images were compressed at a rate of 7 frame/s. Thus, each second of video corresponds to about 21 seconds of measurements.

\noindent\textbf{Mov3-seq3-v1} :
R-PE protein concentration of 2.9 micro Moles (protein clusters are observed). Each 128 x 128 pixel confocal image correspond to a 30 x 30 $\mu$m  field of view and recorded at a pixel dwell time of 0.1 ms at a rate of 2.8 s/frame. Each stack of images was made of 375 images where the laser power has been adjusted every 75 images to 50, 150, 50, 150 and then 50 $\mu$W. An OD1 filter (Thorlabs ND-10B) has been added before the detector. Then, the video made from these images were compressed at a rate of 7 frame/s. Thus, each second of video corresponds to about 21 seconds of measurements.

\noindent\textbf{Mov4-seq4-v1} :
R-PE protein concentration 6.5 micro Moles (protein clusters are observed). Each 128 x 128 pixel confocal image correspond to a 30 x 30 $\mu$m  field of view and recorded at a pixel dwell time of 0.1 ms at a rate of 2.8 s/frame. Each stack of images was made of 375 images where the laser power has been adjusted every 75 images to 50, 150, 50, 150 and then 50 $\mu$W. An OD1 filter (Thorlabs ND-10B) has been added before the detector. Then, the video made from these images were compressed at a rate of 7 frame/s. Thus, each second of video corresponds to about 21 seconds of measurements.

\subsection*{THz spectroscopy experiments}
Long-range interactions between macromolecules can be highlighted through the dependence of their collective  oscillation frequency, lying within the THz domain, on intermolecular distance or concentration. THz near-field spectroscopy allows to measure the occurrence of such collective oscillations if one is able to realize this spectroscopy on proteins immersed in saline solution: a well-known technological roadblock, due to the huge absorption of THz radiations by water. We thus have developed two specific setups to overcome this limitation operating in the 0.07-0.11 THz domain and in the 0.25-0.37 THz range, respectively.

 \textit{\textbf{1. Rectenna-based THz-spectroscopy.}} \par
\begin{figure}[h!]
\centering
\includegraphics[scale=1.7,keepaspectratio=true]{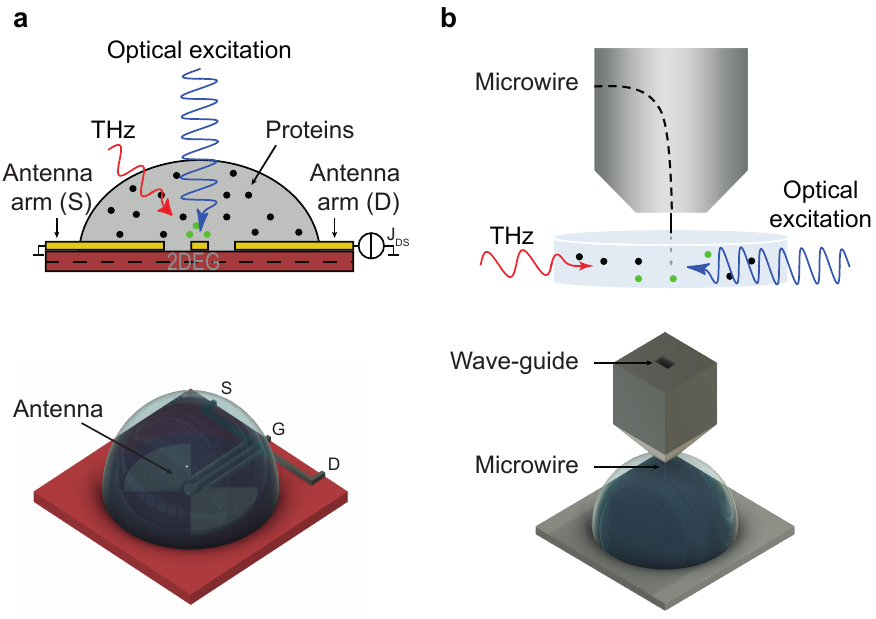} 
\caption{\textbf{THz-spectroscopy experimental setups.} (a) Rectenna for the 0.07 to 0.11 THz band; (b) Microwire-based probe for the 0.25 to 0.37 THz band. }
\label{THz_spectro}
\end{figure}

 
 \noindent\textbf{Setup.} For the 0.07 to 0.11 THz band, a continuous-wave (CW)  THz radiation was generated by a Virginia Diodes WR10 source with 25 mW output power. 
A droplet of 30 - 35 $\mu$L of a solution of R-PE (see Methods) was filed into a 14-pin ceramic dual in-line package (DIL) forming a cuvette where the rectenna (a bow-tie antenna connected to a GaAs/AlGaAs high-electron-mobility transistor)  is placed {(Figure \ref{THz_spectro}(a))}. A plastic-(transparent to blue and THz radiations) film was used to seal the system  and to reduce evaporation. To excite the proteins, a Spectra Physics$^{TM}$ blue laser (ExcelsiorOne 488C-50) emitting at 488 nm and delivering a maximum of 50 mW 
{ (considering a 1 mm beam-diameter, with an optical-power density of $\approx$ 6.4 $W cm^{-2}$) was collimated onto the detector by using a dielectric plane-mirror}. The laser output-power is controlled by a set of optical-density filters. \par
\textbf{Data acquisition.} THz passing through the solution of water, NaCl and excited protein is detected by the rectenna  providing a DC-voltage at its loads, proportional to the THz-field intensity. While the dimensions of the rectenna reach half-a-wavelength (at 0.3 THz), the THz-field is only enhanced in the feed-gap region of the antenna e.g. within a volume of about 0.2 pL. A hydrophobic, biocompatible and transparent to both THz and blue radiations varnish was applied to the upper layer of the rectenna. Experiments were done at room temperature. \par
\textbf{Data analysis.} Normalization procedure, according to Beer--Lambert law, divides spectra of proteins in solution under illumination $I_{R-PE_{ON}}$ by spectra of proteins in solution without illumination $I_{R-PE_{OFF}}$. To obtain spectra of excited proteins only -- $A$ is thus proportional to the absorbance coefficient of excited proteins -- the result is then divided by the buffer spectra $I_{NaCl}$ (also with and without illumination).
\begin{equation}\label{beerlambert}
 A = \log\left( \frac{I_{R-PE_{ON}}\div I_{R-PE_{OFF}}}{I_{NaCl_{ON}}\div I_{NaCl_{OFF}}}\right)\ . 
\end{equation}
Savitsky -- Golay smoothing was used at the end to smooth curves, with 15 points of window and a polynomial order of 2 {(see an example of data treatment in Supplementary Material Figure S9 \cite{cite_supp})}. \par
\textbf{Technical limitations.}
The absence of the second peak in Figure 2 (and in the related Figure S9) is due to the difficulty of working at high input power and long acquisition time, as used for Figure 3. In fact, this would make hard the stabilization of temperature and the control of evaporation (thus of protein concentration), over such a small solution volume and during an overall acquisition time of about 90 minutes that would be needed to observe also the second peak in Figure 2 (instead of 8 - 9 minutes to work it out in its present form, each spectrum having been recorded in 1.6 minutes as is the case of Figure S9). 

 \textit{\textbf{2. Microwire-based probe}}.\par
 \textbf{Setup.} For the 0.25 to 0.37 THz band, CW-THz radiation is produced by a Signal Generator Extension (SGX) module designed by Virginia Diodes, Inc. Module WR9.0 (82-125 GHz) SGX was used with external multiplier WR2.8 to extend the frequency coverage to 0.25-0.37 THz with an average power of 1 mW. The radiation emitted is then collimated and focused by PTFE lenses on the biological sample. Optical excitation, represented by a blue laser and a UV lamp, was used to excite different components of the protein under study. Thus, for the optical excitation of fluorochromes, a Cobolt 06-MLD laser was used emitting an output-power of 40 mW at the wavelength of 488 nm. Optical excitation of aromatic amino acids included in the protein was performed using LED M275L4 produced by ThorLabs at a wavelength of 275 nm and a typical output-power of 60 mW. Due to its large numerical aperture (NA=0.86) UV-coated lenses were used to collimate and focus the radiation onto the sample. Thus, a spot size of about 1cm in diameter was achieved. A micro-coaxial near-field wire (Figure  S\ref{THz_spectro}(b)) is inserted inside a metallic rectangular waveguide to allow a modal transition from TM01 Sommerfeld's mode to TE01 waveguide mode. The probe is capable of highly localized detection of the longitudinal component of the electric field \cite{Keilmann1995:a} due to the sub-wave diameter of the wire around 10 $\mu$m at the tip into the volume of about 4 pL.\par 

\textbf{Data acquisition.} A specific sample holder was also designed. It is constituted by a plastic cylinder, transparent to UV, blue and THz radiations and resistant to UV radiation with a diameter of 7 mm and a volume of 53$\ \mu$L. At its bottom, a 130$\ \mu$m-thick glass was used, through which laser excitation was performed. To avoid evaporation and, as a consequence, changes in protein concentration during the experiment, the sample was covered  by a plastic film with a thickness of 5$\ \mu$m, allowing the wire to pass through it without damage.  \par

\textbf{Data analysis.} Regarding the normalization, the same procedure than in rectenna experiments has been followed, according to Beer--Lambert law, to obtain spectra of excited proteins only. However, since in this frequency-band (0.25 - 0.37 THz) the signal-to-noise ratio is poor due to high absorption by water molecules, a specific data processing and normalization procedure (using a band-stop Fourier transform filtering and Lorentz functions fitting)  has been followed to extract an usable signal and frequencies of collective-oscillation resonances. 

It is thus possible to observe the evolution of the absorption-peak amplitude as a function of the duration of illumination. Were considered as absorption peaks, those whose height and area are increasing continuously with time. Time (up to 1 hour) and concentration (from 750 to 60,000 nM) dependences of the THz-spectra have been characterized. 
\nocite{richards1977areas}
\nocite{durchschlag1989determination}
\nocite{chavanis2002phase}
\nocite{likhachev2003dependence}
\nocite{israelachvili2011intermolecular}
\nocite{farnum1999effect}
\nocite{lund2003mesoscopic}
\nocite{zhang2019quantum}
\nocite{faraji2020exciting}
\nocite{torshin2001charge}
\nocite{tsonchev2004screened}
\nocite{alberts2002molecular}
\nocite{zaragoza2019detecting}
\nocite{campa2009statistical}
\nocite{shcherbakov2014dielectric}


\bibliographystyle{naturemag}
\bibliography{scibib}

\section*{Acknowledgements}
The project leading to this publication has received funding from the Excellence Initiative of Aix-Marseille University - A*MIDEX, a French "Investissements d'Avenir" programme. This work was also partially supported by the Seventh Framework Programme for Research of the European Commission under FET-Proactive grant TOPDRIM (FP7-ICT-318121), by the projects SIDERANT and NEBULA financed by the french CNRS, by the Occitanie Region and by Montpellier University through its TOP platform and by the LabEx NUMEV (ANR-10-LABX-0020) within the I-Site MUSE. We acknowledge the PICSL imaging facility of the CIML (ImagImm), member of the national infrastructure France-BioImaging supported by the French National Research Agency (ANR-10-INBS-04).This project has also received funding from the European Union Horizon 2020 Research and Innovation Programme under the Marie Sk{\l}odowska-Curie grant agreement No 765426 (TeraApps). 
This project has received funding from the European Union's Horizon 2020 research and innovation programme under grant agreement No 964203 (FET-Open LINkS project).
 
\medskip

\subsection*{Authors Contributions}
M.L., Y.M., M.G. have been crucial for the success of the project. M.L. designed and performed all the experiments in Marseille regarding diffusion, he fixed the problem of operating the FCCS at high concentrations, and made the successful choice of the proteins to be studied. M.G. performed all the theoretical modelling and the related Molecular Dynamics and Monte Carlo numerical computations. Y.M. and  S.R. carried out the experiments in Montpellier concerning THz-spectroscopy using the rectenna sensor and the micro-wire probe, both used for intramolecular collective vibration and frequency shift measurements. I.N. participated in the project from its very beginning many years ago paving the way to the present findings and suggested that ED forces could be detected through the frequency shift reported here. E.F. participated in the project from its very beginning and contributed the theoretical part of the paper. S.M. and D.M. gave fundamental support to perform the FCS and FCCS experiments. P.F., F.T. and L.V. participated in the project supporting it since its very beginning many years ago. J.S. contributed on the biophysical and biochemical aspects of the work. J.T., with the support of D.C. and S.R., conceived, designed and built the experimental setup in Montpellier and supervised the experiments. M.P., in quality of project leader, initiated this research many years ago and designed, supervised and intervened in all the theoretical and experimental aspects of the project. All the authors contributed to the discussion and to the analysis of the results. M.P. wrote the paper with the help of J.T.,M.G., D.M. and J.S.

\subsection*{Data Availability}
Additional data that support the findings of this study are available in the online version of the paper as Supplementary Materials. Videos showing the formation of clusters are also available as online materials.

\subsection*{Code Availability}
All the details to reproduce the numerical implementation of the theoretical models to interpret the experimental outcomes are available in the online version of the paper as Supplementary Materials.

\subsection*{Corresponding author}
Correspondence and requests for materials should be addressed to M.P. and J.T.

\subsection*{Competing interests}
The authors declare no competing interests.

\subsection*{Supplementary Materials}

Additional Experimental Data

Theoretical interpretation of the experimental results

Fig S1 - S22

\textbf{Online also:}

Movies S1 - S4

All this material can be found at  https://www.science.org/doi/10.1126/sciadv.abl5855

\end{document}